# Noise Induced Synchronization on Collective Dynamics of Citrus Production

Nina Sviridova [1] and Kenshi Sakai [2][†]

**Abstract**

It is very common to observe nonlinear features in agricultural and ecological systems. For example, in tree crop production, alternate bearing is well known phenomena caused by nonlinear dynamics. Production of single tree of *Citrus Unshiu* trees is recognized to be driven by a mechanistic process modeled with the so-called "resource budget model", which demonstrates phenomenon of alternate bearing. However, the term of alternative bearing is used not only for an individual tree's production but also for total production of a large sized population of trees. In this paper, we developed noise induced uncoupled dynamics model for population alternate bearing based on Isagi's resource budget model. Based on numerical experiments with the developed model, theoretical possibility of a substantial alternate bearing effect even in a national market was proposed.

[Keyword] alternate bearing, nonlinear dynamics, resource budget model, bifurcation, diagram, noise induced synchronization

## 1. Introduction

Toward innovative research and developments in the production system of citrus and nuts industry, it is very important for bio-system engineering to understand fundamental principles of alternating bearing and masting. Various species of tree crops demonstrate significant fluctuations of annual yield: producing large amount of flowers and fruits at one year (on-year) and significantly smaller at following years (off-year) until next on-year. This phenomenon is well known as "alternate bearing" for citrus and "masting" for nuts and acorn.

The term of alternative bearing can be used not only for an individual citrus tree's production but also for total production of population from orchard to national market level. Alternate bearing of *Citrus Unshiu* has occurred clearly in the national market level and leading large fluctuations of market production and price in Japan. Therefore, alternate bearing and masting have been investigated in two dimensions such as individual level and population level. It has been one of the most important research topics in pomology to develop methods to control the alternate bearing of individual tree. Additionally, the dynamics of national market productions and pricing of *Citrus Unshiu* has been an attractive problem in agricultural economics.

The so-called "resource budget model (RBM)" (Isagi et al., 1997) was developed to explain a mechanistic process of acorn masting in individual tree level. Sakai (2001) employed RBM to explain the individual level alternate bearing of *Citrus Unshiu* and validated the dynamics expected by RBM with experimental data (Noguchi, 2003; Sakai et al., 2008).

In order to explain population level acorn masting, "pollen coupling" mechanism was incorporated for cross-pollinated types acorn trees (Isagi et al., 1997; Satake and Iwasa, 2000). In the expanded RBM, the masting is recognized as a synchrony of population. However, in case of citrus we are not to able assume coupling mechanism in RBM to expand for alternate bearing in population, because *Citrus Unshiu* is self-pollinated plant. As the number of citrus trees in Japan is estimated at the order of 100 million, the large fluctuations of individual trees should be smoothed out and significant fluctuations of total annual production should not be expected without any rigorous explanation.

In population ecology, "Moran effect" is known as population synchrony caused by climatic forces (Moran, 1953; Royama, 1992; Satake and Iwasa, 2002). Common noise induced synchrony is also well known phenomena as a group of nonlinear oscillators can synchronize when the common environmental noise is induced (Kurebayashi et al., 2012) on each oscillator. As real systems are always containing noise attributed to external random forces, it seems to be reasonable to set a hypothesis of environmental condition fluctuation effect into RBM. However, without coupling term in RBM, noise induction was recognized to be insufficient to explain alternate bearing as a synchrony in population (Satake and Iwasa, 2000).

This study sought to investigate the effect of common noise induced RBM as one of possible mechanisms explaining Moran's effect. In other words, the main purpose of this paper is to develop a model which can explain the alternate bearing in a large size population including national production level.

---
[1] Student Member: United Graduate School of Agriculture, Tokyo University of Agriculture and Technology, 3-5-8 Saiwai-cho, Fuchu-shi, Tokyo 183-8509, Japan;
Present: Computing Center, Far East Brach Russian Academy of Science, 65,Kim-Yu-Chen, Khabarovsk, 680000, Russia
[2] Member: Corresponding author, Environmental and Agricultural Engineering Department, Tokyo University of Agriculture and Technology 3-5-8 Saiwai-cho, Fuchu-shi, Tokyo 183-8509, Japan;
[†] Corresponding Author: ken@cc.tuat.ac.jp

## 2. Alternate bearing of Citrus Unshiu in terms of individual and national production

At the Nebukawa Experimental station in Kanagawa Prefecture, the number of fruits from two individual trees that were not subjected to any conventional operations such as pruning and branching were measured for 48 years. Very clear alternate bearing were observed for them as shown in Fig. 1. The phase of these two motions was completely opposite with implying independency between individual trees. This is the example of alternate bearing for Citrus Unshiu in individual level. On the other hand, the term of alternate bearing is also used for describing total production of citrus fruits in a large size of population. For this example, the total production of Citrus Unshiu in Japan is shown in Fig. 2(a). The trend of decline is obvious because of orange and other citrus fruits imported from outside countries. In order to eliminate the trend caused by international trade, the increment of the production is shown in Fig. 2(b). It is very clear that a consistent On-Off pattern exists in it. This pattern is the typical example of alternate bearing for total production of citrus fruits in a large size of population.

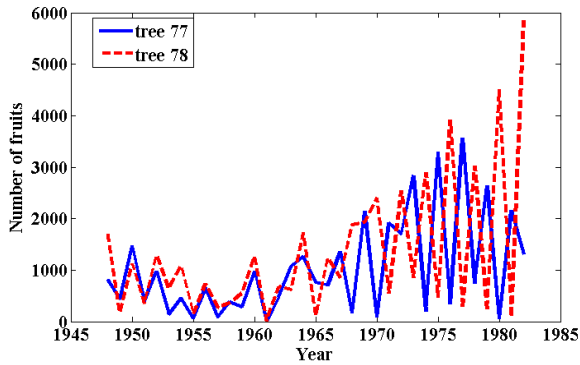

Fig.1 Alternate bearing of two individual trees

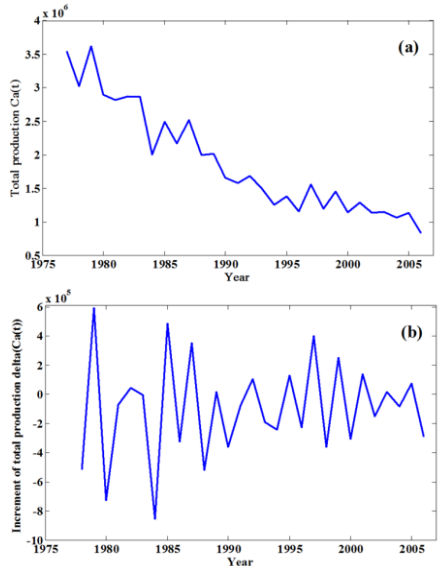

Fig2 Census data of Citrus Unshiu in Japan from 1976 to 2006: (a) National production; (b) Increment of the national production

## 3. Common noise induced RBM for N trees

The model we are developing to explain the alternative bearing on total production of citrus in a large size population is based on RBM for an individual plant (Isagi et al., 1997; Sakai, 2001). This conventional model assumes that there is a constant amount of photosynthate produced by an individual plant during a year. The photosynthate is used for growth and maintenance of the plant, and surplus $P_S$ accumulates at the plant body's trunk. Let $I(t)$ be the amount of energy reserve at the beginning of year $t$. When accumulated photosynthate ($I(t)+P_S$) exceeds the threshold of the pool $L_t$, the excess amount ($I(t) +P_S – L_T$) is used as the cost of flowering $C_f$. The cost of pollinating flowers and bearing fruits is designated as $C_a$. The ratio $C_a/C_f$ is assumed to be a constant, $R_C$. After the reproductive stage, the accumulated photosynthate becomes

$$L_T - C_a = L_T - R_C C_f.$$

Surplus photosynthate $P_S$ in the conventional RBM is assumed to be constant, however natural systems are usually subjected to various external random forces such as temperature changes (Satake & Iwasa,2000; Satake & Iwasa, 2002), that is assumed to appear in form of factors that results in $P_S$ fluctuation. Our common noise induced RBM for a population which size is $N$ is described by equations (1)-(3). In this model, we employed pulse noise with constant amplitude $\alpha$ and pulse interval of noise is given as uniform random distribution. The noise is induced on $P_S$ as shown in the right side of equations (1) and (2). Here, it should be noted that the noise is induced identically on each individual tree. In this study noise amplitude $\alpha$ was empirically chosen $0.05$ which as seen in (1)-(2) induces 5 % pulse noise on $P_S$. Conventional RBM can be obtained from equations (1)-(3) by choosing $\alpha=0$, in this case stochastic component inducing noise on $P_S$ will be eliminated. As seen in system (1)-(3) the developed model does not include coupling component.

$$I^i(t+1) = \begin{cases} I^i(t) + P_s[1 + \alpha H(\theta(t) - h)], & \text{if } I^i(t) + P_s \le L_T \\ I^i(t) + P_s[1 + \alpha H(\theta(t) - h)] - C_f^i(t) - C_a^i(t), & \text{if } I^i(t) + P_s > L_T \end{cases} \quad (1)$$

$$C_f^i(t) = I^i(t) + P_s[1 + \alpha H(\theta(t) - h)] - L_T, \quad (2)$$

$$C_a^i(t) = R_C C_f^i(t), \quad (3)$$

where $I^i(t)$ is the amount of energy reserve at the beginning of year $t$ for tree $i$; $C_a^i(t)$ is the quantity of fruits produced by tree $i$ at the end of year $t$; $C_f^i(t)$ is the cost of flowering for tree $i$ in year $t$; the system parameters $P_S$, $L_T$, and $R_C$ are assumed to be the same for all citrus trees. $H(x)$ is the Heaviside step function defined as

$$H(x) = \begin{cases} 0, & \text{if } x < 0 \\ 1, & \text{if } x \ge 0 \end{cases}$$

and $\theta$ is vector with uniform random components, where $\theta(t) \in [0; 1]$; $h \in (0; 1)$ is a constant.

In this model, the total citrus production can be defined as the sum of fruits $C_a^i(t)$ ($i=1..N$) from each single tree.

Let $Av(t)$ be the averaged production of one tree in a population consisting of $N$ trees in year $t$:

$$Av(t) = \frac{1}{N}\sum_{i=1}^{N} C_a^i(t). \qquad (4)$$

Since as seen in (4) $Av(t)$ and the total production have the same dynamics, we use $Av(t)$ as index of behavior of total production in the following discussion.

## 4. Numerical experiments

In order to investigate the performance of the developed model for describing the alternate bearing in a large sized population, we conducted two types of numerical experiments. One is conducted for the effect of pulse noise on the behavior of production of single tree and the other is for total production of $N$ trees.

### 4.1. Production of singe tree

As it was mentioned above, single tree model RBM without coupling and noise that can be expressed by equations (1)-(3) with $\alpha=0$ demonstrates alternate bearing. Fig. 3(a) and Fig. 3(b) show the time series of $C_a$ on noise free RBM and noise induced RBM, respectively. Percentage of pulse noise randomly induced on $P_S$ is shown in Fig. 3(c), as it can be seen according to pulse noise component introduced into the RBM model and choice parameter $\alpha=0.05$ for each year two scenario is possible either 5 % noise is induced on $P_S$, or there is no noise induction. RBM parameters are set as $P_S=10$, $L_T=100$ and $R_C=2.5$, which value was discussed by Isagi et al. (1997). As seen in Fig. 3(a) and Fig. 3(b) there is no significant difference observed between two time series. In order to investigate the global behavior over $R_C$, the bifurcation diagrams on noise free and noise induced RBM are shown in Fig. 4(a) and Fig. 4(b), respectively. It is also difficult find out the significant effect of induced noise on RBM as long as single tree production is observed.

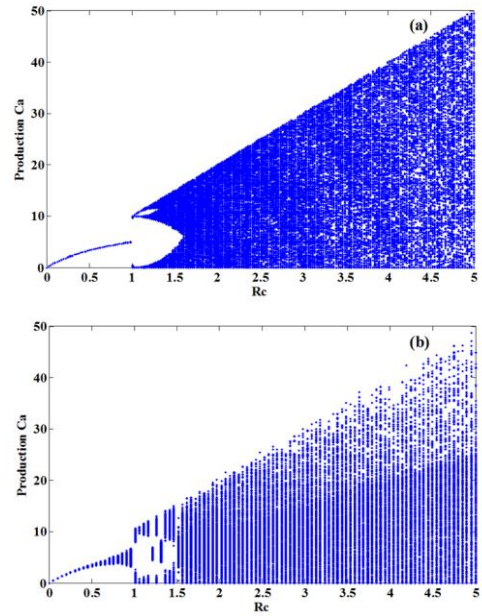

Fig.4 Bifurcation diagrams of alternate bearing on individual tree: (a) bifurcation diagram for noise free RBM; (b) bifurcation diagram for noise induced RBM

### 4.2. Total production of N trees

Instead of observing production of single tree, the behavior of total production of $N$ trees are investigated in Fig. 5. In the numerical experiments, population size $N$ was set as 100,000. The time series of $Av(t)$ for noise free and noise induced RBM with $R_C=2.5$ are plotted in Fig. 5(a) and Fig. 5(b), respectively. As seen in Fig. 5(a), $Av(t)$ is almost a constant, in the contrary, a significant On-Off fluctuations are observed in noise induced RBM in Fig. 5(b). The induced pulse noise that was chosen same as in case of single tree is shown in the Fig. 5(c).

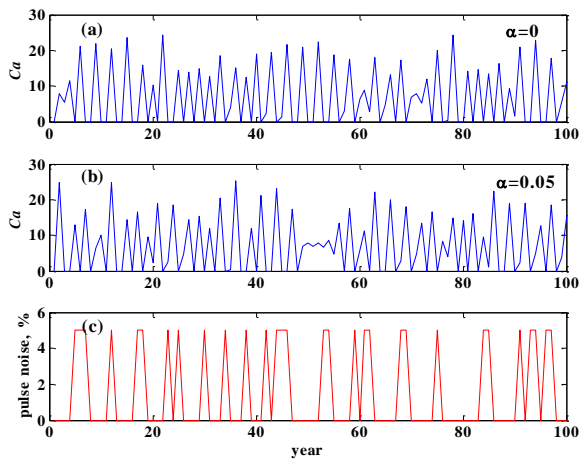

Fig.3 Dynamics of alternate bearing on individual tree: (a) production $C_a$ for noise free RBM ($\alpha=0$); (b) production $C_a$ for RBM with 5 % noise induced on $P_S$ ($\alpha=0.05$); (c) induced pulse noise

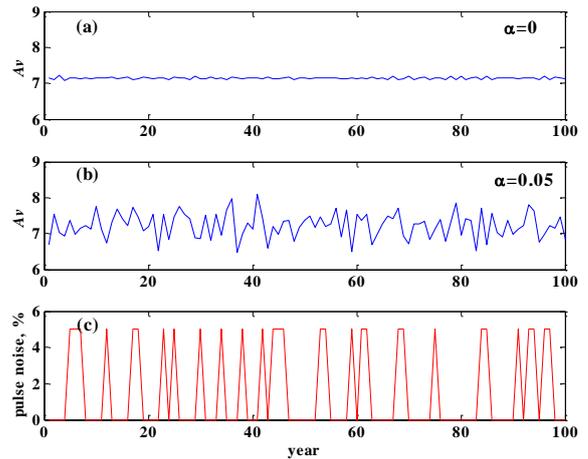

Fig.5 Dynamics of alternate bearing in a large sized population consisting of 100,000 trees: (a) averaged production $Av$ for noise free RBM ($\alpha=0$); (b) averaged production $Av$ for RBM with 5 % noise induced on $P_S$ ($\alpha=0.05$); (c) induced pulse noise

By changing $R_C$ from 0 to 5, the bifurcation diagrams of $Av(t)$ are shown in Fig. 6(a) for noise free and Fig. 6(b) for pulse noise induced RBM. At each $R_C$, 100 values of $Av(t)$ corresponding to $t= 101$ to 200 are plotted on the bifurcation diagram. For all the range of $R_C$, fluctuations of $Av(t)$ for noise free RBM are very small, however in case of noise induction, significant large fluctuations, which can be recognized as alternate bearing in the population, are clearly observed. Additionally, we introduced coefficient of variance for $Av(t)$ ($t= 101$ to 200) as following:

$$CV_{av}(t) = \frac{\sqrt{\frac{1}{100}\sum_{t=101}^{200}(Av(t)-MA)^2}}{MA} \quad (5)$$

where, $MA = \frac{1}{100}\sum_{t=101}^{t=200} Av(t)$.

$CV_{av}(t)$ in (5) is defined at a given $R_C$ and is used to characterizes changes in production variation with increasing the population size $N$. Fig 7 reflects $CV_{av}(t)$ changes for $N=1000$ to 300,000 (1,000, 10,000, 100,000, 300,000) in logarithmic scale at $R_C=5$. As it can be seen for noise free model variation ($CV_{av}$) of total production tend to decrease toward zero, while for 5 % pulse noise induced RBM it stabilizes in a constant value with small fluctuations. This result suggests that noise induction in RBM causes presence of annual total production fluctuations (i.e. alternate bearing) that cannot be observed for noise free model.

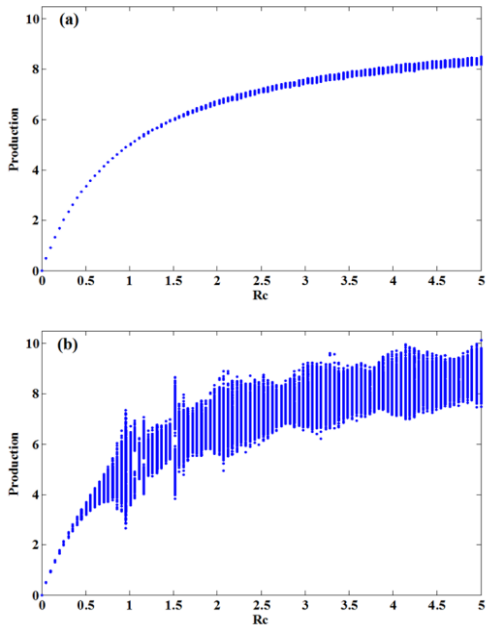

Fig.6 Dynamics of alternate bearing in a large sized population: (a) bifurcation diagram for noise free RBM; (b) bifurcation diagram for noise induced RBM

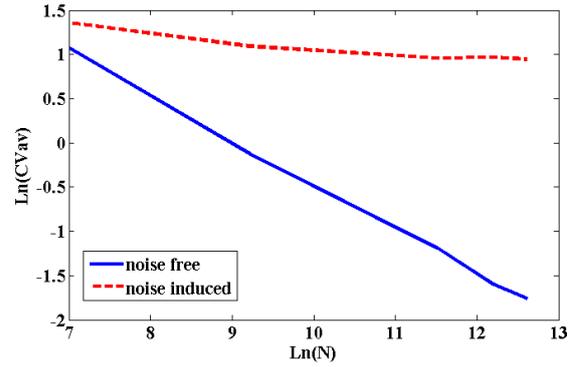

Fig.7 The effect of population size $N$ on alternate bearing on a population at $R_C=5$

## 5. Discussion

Large fluctuations of fruits and seeds of tree crops and forest trees have been investigated in citrus, nuts and acorn as alternate bearing or masting phenomena (Kelly et al., 2002; Rosenstock et al., 2011; Koenig et al., 2003). In theoretical population ecology, masting of acorn is modelled as coupled dynamics consisting of nonlinear oscillators described by RBM (Isagi et al., 1997; Iwasa & Satake, 2000; Lyles et al., 2009; Akita et al., 2008). As most masting acorn trees show cross-pollination, the pollen coupling can be a key factor to generate synchrony among trees which compose the population. However, *Citrus Unshiu* is a self-pollination plant, therefore we can't take into account pollen coupling as a factor to generate synchrony.

Synchronization of annual yield of large population of trees has been subject of numerous studies and several explanations were proposed at different times. Some of them, such as hypothesis of environmental condition fluctuation effect, were later recognized to be insufficient to explain masting in large group of trees (Satake and Iwasa, 2000). In our model, we employed pulse noise for modelling external natural random effects, such as whether cue, that are empirically known as a source of alternate bearing or masting. Additionally, we focus on the variance of total production and do not evaluate their synchrony with correlation among trees. In developed noise induce RBM the alternate bearing in a large sized population can be observed, additionally our noise induced RBM allows investigation of total production behavior. Following investigation of the developed model in terms of noise induced un-coupled dynamics or stochastic resonance can be fruitful for bio-system engineering.

## 6. Conclusions

In this study, the theoretical possibility of existence of alternate bearing in total production on a level of national market was demonstrated by proposing noise induced RBM for $N$ trees. For modelling whether cue such as for example temperature for summer growing season, noise is induced on the rest of the photosynthate $P_S$ of RBM. The induced noise is given as random pulse whose amplitude is constant (5 % of $P_S$) and time periods are randomized. The developed model demonstrates clear On-Off patterns simulating alternate bearing on total production. The effect of inducing noise is getting clearer with increasing of number of trees $N$. Obtained results clearly showed that with independent fruiting (without coupling) even for enough large number of trees the total production will demonstrate alternate bearing. Proposed model can make a possible explanation for alternate bearing phenomenon observed on national citrus market in Japan. The discovered results in this paper for explaining population level alternate bearing have potential to supply a fundamental principal to predict national level citrus production.

## Acknowledgments

This work was supported by JSPS Grant-in-Aid No.25660204. N.S. thanks MEXT for providing the Japanese Government Scholarship for three years. Authors also would like to express their gratitude to anonymous reviewers whose comments helped improve this manuscript.

温州ミカンの集団力学系における共通ノイズ同期
ニーナ・スヴィリドヴァ[*1], 酒井憲司[*2†]
要旨
農業生態系において非線形現象は遍在的である。例えば、柑橘類における隔年結果現象は非線形ダイナミクスの典型例として知られている。温州ミカンの樹木個体レベルで隔年結果現象は、いわゆる資源収支モデルで記述されるメカニズムによって説明されている。一方、隔年結果という用語は、樹木個体レベルの収量変動の現象だけではなく、温州ミカンの国内市場における収量や価格の変動現象に対しても用いられる。本研究では、個体レベルの変動モデルを振動子と考えて、大量の振動子集団に共通ノイズを加えることによって、振動子間に結合を有しない場合にも集団レベルでの同調現象が隔年結果とし生じることを数値実験によって明らかにした。

[キーワード] 隔年結果, 非線形力学,物質収支モデル, 分岐図, ノイズ印加同期

1.学生会員:東京農工大学連合農学研究科，（〒183-8509　東京都府中市幸町 3-85-8　TEL 042-367-5755）

現在: ロシア科学アカデミー極東支部計算センター，（65,Kim-Yu-Chen, Khabarovsk, 680000, Russia）

2. 會員:責任著者、東京農工大学農学研究院、（〒183-8509　東京都府中市幸町 3-85-8 TEL 042-367-5755）